\documentclass[12pt]{article}
\usepackage{graphicx}  
\begin{document}
\baselineskip=24pt
\title{A study of frequency and size distribution dependence of extinction 
for astronomical silicate and graphite grains}
\author{Ashim K. Roy$^{1}$\thanks{E-mail:ashim@isical.ac.in (AKR); 
sharma@ose.res.in (SKS); rag@iucaa.ernet.in (RG) } Subodh 
 K. Sharma$^{2}$ and Ranjan Gupta$^3$\\\\
$^{1}$Indian Statistical Institute, 203, BT Road, Kolkata 700108, India\\
$^{2}$ S N Bose National Centre for Basic Sciences, Kolkata 700098, India\\
$^{3}$ Inter University Centre for Astronomy and Astrophysics, Pune 411007,  
India}

\maketitle
\clearpage
\newpage
\noindent
\newpage
\label{firstpage}

\begin{abstract}
It is generally agreed that interstellar dust grains consist of two main 
components, namely, silicates and graphites. Some models, like MRN model, 
assume these grains to be homogeneous spheres following a power law size 
distribution. This paper presents, in the framework of Mie theory, a 
parametrization of extinction spectrum curves of the silicates and the 
graphites separately in terms of frequency and the minimum and maximum of  
sizes in the distribution. Analytic expressions in  
ultraviolet and far-ultraviolet are presented for both types of grains.The 
values of maximum and minimum sizes for which these equations are valid have 
been identified. These   
equations can be useful in a number of situations involving silicate and 
graphite grains.  
\end{abstract}

%

\newpage
\noindent

\section{Introduction}

 In a recent publication we analyzed extinction spectrum of a collection of 
homogeneous spherical particles of a unknown size distribution [1] ( 
hereafter referred as RS-I). It was shown that an extinction spectrum, in 
general, has some easily identifiable characteristic regions where the 
extinction-frequency relationship can be approximated by simple empirical 
formulae involving first four moments of the particle size distribution. The 
analysis clearly exhibited the manner in which  essential features of the 
particle size distribution gets coded into its extinction spectrum (more
generally known as the Interstellar Extinction Curve in the astrophysical
situation). It was 
demonstrated that the moments could indeed be obtained from its extinction 
spectrum and that with the knowledge of these moments it was possible to 
reconstruct the size distribution function. \\\\
The analysis in RS-I however, assumed that all particles were of same material  
for which the refractive index did not vary with frequency. The question 
arises what happens when a collection of particles has more than one material 
component with their refractive indices varying with wavelength. 
 We address ourselves to these questions in this 
paper. In other words, the purpose of 
this work is to examine the possibility of extending the  ideas developed in 
RS-I to a multi-component system where refractive index of the individual components
 varies with wavelength of incident electromagnetic radiation. \\\\ 
 One suitable model where these extensions of RS-I can be studied is a modified
version of interstellar dust model due to Mathis, Rumpl and Nordsieck (MRN) 
[2]. The MRN model assumes that interstellar dust consists of individual grains
 of homogeneous spheres of silicates and graphites having very definite   
 power-law size 
distributions. For graphite grains,  there is a ``$\frac{1}{3}-\frac{2}{3}$" 
approximation [3]. This modification essentially converts two component MRN 
model to a three component MRN model. Investigations in this work show that 
 the frequency-size relationship can still be obtained even when the refractive
 index of the grains is varying with wavelength. But, because of rapid variation
 of refractive index with frequency, now one needs to divide the spectrum in 
suitable intervals and obtain separate relations for each sub-interval. This 
paper presents these relationships in ultraviolet and far-ultraviolet frequency
 domains for silicate and graphite grains individually. Thus, we not only 
demonstrate, the successful extension of ideas in RS-I to materials whose 
refractive index varies with wavelength, but in the process we also present 
relationships which can be used for testing various models of interstellar dust.\\\\  
This paper is organized as follows. Section 2 describes  the dust model 
considered here. Section 3 gives the functional form of extinction in terms of  
size distribution parameters as well as the frequency for both graphite and 
silicate grains. Subsequently, numerical results of extinction obtained from these formulae 
are compared with exact Mie theory computations in section 3. The possible use 
of such formulae for individual components to multi-component system has been 
demonstrated in section 4. Finally, we conclude by 
summarizing and discussing the results of this paper in section 5.
\section{The Dust Model} 
A classic model of interstellar dust was proposed more than 30 years ago by  
Mathis, Rumpl and Nordsieck (MRN) (1977). Since then, the basic  model is 
being used even today albeit  
with some modifications. It has not been fully superseded by later studies. 
The MRN model uses two separate populations of bare silicate and graphite 
grains with a power-law distribution of sizes of the form:  
\begin{eqnarray}
f(a) \propto a^{-3.5}~~~~~~a_0\leq a \leq a_m, 
\label{psd}
\end{eqnarray} 
 where $a$ denotes the radius of the spherical grain varying within the limits
 $a_0$ (minimum radius) and $a_m$ (maximum radius). The size range for  
graphite and silicate grains is:
\begin{eqnarray} 
 {Graphite~ grains}:~~~~a_0\sim 0.005\mu m~~~~a_m\sim 0.25\mu m, \\
 {Silicate~ grains}:~~~~a_0\sim 0.005\mu m~~~~a_m\sim 0.25\mu m.
\label{range}
\end{eqnarray}
Further, the graphite material is supposed to be present in two distinct 
structural varieties within the specified range of $a_0$ and $a_m$. This 
plausibility lies in the fact that graphite is highly anisotropic 
material. The refractive index of graphite, 
therefore, depends on the orientation of electric field relative to the 
 structural symmetry. Owing to difficulties in calculations of exact 
scattering quantities 
due to anisotropy, workers have taken resort to an 
approximation known as $``\frac{1}{3}-\frac{2}{3}"$ approximation [3]. In this  
approximation graphite
 grains are represented as mixture of isotropic spheres, $\frac{1}{3}$ of which
 have refractive index $m=m_{\parallel}$ (referred as graphite parallel) and 
$\frac{2}{3}$ have the refractive 
index $m=m_{\perp}$ (referred as graphite perpendicular). This modification effectively makes MRN model a three 
component model.
\section{Extinction as a function of frequency and size distribution 
parameters}
Exact extinction coefficient, $K_{ext}, $
 was obtained using the formula,
\begin{eqnarray}
K_{ext}(\lambda)=\pi N\int_{a_0}^{a_m} Q_{ext}(x)a^2 f(a)~ da, 
\label{extinction}
\end{eqnarray}
where $Q_{ext}(x)$ is the extinction efficiency of an individual scatterer of 
size parameter $x$ and  $x=2\pi a/\lambda$ with 
$\lambda$ as the wavelength of the radiation. The  
exact extinction efficiency $Q_{ext}$ for a  spherical homogeneous 
scatterer can be computed using Mie formulas.
 The necessary refractive index particulars for various components (at various 
wavelengths) were taken from the tables provided by Draine [4] on his website. 
The number of particles per unit volume, $N$ has been arbitrarily fixed at 
$N= 4.4\times 10^8$.
 in these calculations. This being a multiplicative constant, its admitted 
value does 
not make any effective difference in the functional form of $K_{ext}$ we wish 
to determine. \\\\
In case of size distributions of spherical particles having constant refractive
 index (wavelength independent) it was shown in RS1 that in the $K_{ext}-\nu$ graph, 
in general, we can identify frequency intervals where $K_{ext}$ has 
distinctly linear or parabolic or asymptotic behaviour having corresponding 
functional forms of $K_{ext}$:
$$
K_{ext}^{(L)} \sim l_1(m)\bar{a^3} \nu + l_2(m) \bar{a^2}, $$
$$
K_{ext}^{(P)} \sim p_1(m)\bar{a^4} \nu^2 + p_2(m) \bar{a^3}\nu +p_3(m)\bar{ a^2}, $$
etc., where, $l_1, l_2, p_1, p_2, p_3$ are coefficients which have arbitrary 
dependence on refractive index $m$. $\bar{a^2}, \bar{a^3}$ and $\bar{a^4}$ 
are the 2nd, 3rd and 4th raw moments of the size distribution $f(a)$.\\\\
In the present situation, where $m$ varies with $\nu$, the coefficients $l_1, 
l_2, p_1, p_2, p_3$ will also show variation with $\nu$ and hence the 
distinctly simple linear and parabolic forms of $K_{ext}$ will not be observed 
within a meaningful frequency interval. Hence, our approach has been to study 
the $K_{ext}-\nu $ graph for each material component separately for various 
ranges of $a$, which means varying $a_0$, $a_m$. In case of the power-law 
size distribution, $f(a)=ca^{-3.5}$, we have relations:
$$c=\frac{5}{2}a_0^{5/2}\Bigl(1-\frac{1}{n^5}\Bigr)^{-1} ,$$
$$\bar {a}=\frac{5}{3}a_0\Bigl(1-\frac{1}{n^3}\Bigr)
\Bigl(1-\frac{1}{n^5}\Bigr)^{-1} ,$$
$$\bar {a^2}=5{a_0^2}\Bigl(1-\frac{1}{n}\Bigr)
\Bigl(1-\frac{1}{n^5}\Bigr)^{-1} ,$$
$$\bar {a^3}=5{a_0^3}\Bigl(n-1\Bigr)
\Bigl(1-\frac{1}{n^5}\Bigr)^{-1} ,$$
$$\bar {a^4}=\frac{5}{3}{a_0^4}\Bigl(n^3-1\Bigr)
\Bigl(1-\frac{1}{n^5}\Bigr)^{-1} ,$$
where $n=(a_m/a_0)^{1/2}$. Consequently, with a functional form $K_{ext}(a_0,
a_m, \nu)$, 
our investigations reveal that the extinction in the UV and FUV regions for the 
materials considered have the following  
general form:
\begin{eqnarray}
K_{ext} =CNa_{0}^{5/2}\Bigl[\phi(a_0, \nu)+\psi(a_{m},\nu)\Bigr]. 
\label{general}
\end{eqnarray}
The functions $\phi$ and $\psi$ have forms which change in various frequency 
sub-intervals. For each component of graphite (parallel and perpendicular), four 
formulae were needed to fit extinction. For silicate, three formulae were 
sufficient. Several values of $a_{m}$ and $a_{0}$ were considered.   
 Formulae presented here are valid in the size limits:
\begin{eqnarray}
 {Graphite~ grains}:~~~~0.002\mu m\leq a_0\leq 0.005\mu m;~~0.15\mu m\leq a_m\leq 0.25\mu m,\\
 {Silicate~ grains}:~~~~0.004\mu m\leq a_{0}\leq 0.006\mu m;~~ 
0.2\mu m\leq a_{m}\leq 0.4\mu m.
\label{range1}
\end{eqnarray} 
Observations of the spectra suggested that the functions $\phi(a_0,\nu)$ and 
$\Psi(a_m,\nu)$ can have the simple forms:
$$\phi(a_0,\nu)=b_0(\nu)+a_0^{1/2}b_1(\nu)+a_0b_2(\nu) $$
$$\Psi(a_m,\nu)=\frac{c_0(\nu)}{a_m^{1/2}}+\frac{c_1(\nu)}{a_m}+\frac{c_2(\nu)}{a_m^2}+ \frac{c_3(\nu)}{a_m^3}+....., $$
where $a_0, a_m$ are taken in units of $10^{-5} cm$, $\nu$ in units of $10^5 cm^{-1}$. The number of significant terms contributing to extinction in $\phi$ and
$\Psi$ depend on the material as well as the frequency interval considered. The
 forms of $\phi(a_0,\nu)$ and $\Psi(a_m,\nu)$ have been constructed by careful 
analysis of the regional (frequency sub intervals) behaviour of the extinction 
for each of the materials considered. In the process of developing analytic formulae, care has been taken to have a good compromise between accuracy and 
calculational simplicity so that our analysis could be applied expediently for 
purposeful dust modelling within the power law framework considered here.\\\\   
Following relationships have been obtained:
\subsection{\bf  Homogeneous graphite grains with refractive index $m=m_{\perp}$ }
{\bf 1. For $1000\leq \lambda \leq 1460$\AA~ (FUV region I)}:
$$K=C a_{0}^{5/2}\Biggl[\nu^2\Big(0.259+20.3073(\nu-0.8485)^4)-(\nu a_0)^{1/2}\Bigl( 0.267\nu-0.16048+$$
$$71.15(\nu-0.8428)^4\Bigr) -a_0\nu^{5/2}\Bigl(0.0458\nu +0.0164+$$
\begin{eqnarray}
14.8274(\nu-0.8428)^2(\nu-0.67905)\Bigr)-
0.0488\Bigl(\frac{1}{a_m^{1/2}}+\frac{0.1551}{a_m}\Bigr)\Biggr]
\label{graphiteperp1}
\end{eqnarray}
{\bf 2. For $1460\leq \lambda \leq 1900$\AA~ (FUV region II)}:
$$K=C a_0^{5/2}\Biggl[1.97754-\frac{2.42703}{\nu}+\frac{0.79483}{\nu^2}-\frac{a_0^{1/2}}{\nu^2}\Bigl(0.2164-0.27812\nu+
\frac{0.43347}{\nu^{1/4}}(1.0-\frac{0.6856}{\nu})$$
$$(1.0-\frac{0.5263}{\nu})\Bigr)-
\frac{a_0}{2\nu^2}\Bigl((1.0-\frac{0.606}{\nu})^2
+\frac{0.03173}{\nu}(1 -\frac{0.6262}{\nu}) - 46.52\nu(\nu -0.6856)$$
\begin{eqnarray}
(\nu-0.5263)(\nu -0.606)\Bigr)
-0.0488(\frac{1}{a_m^{1/2}}+\frac{0.1551}{a_m})(0.8667+24.1(\nu-0.6112)^2)\Biggr].
\label{graphiteperp2}
\end{eqnarray}
{\bf 3. For $1900\leq \lambda \leq 2500$\AA~ (UV region I)}:
$$K=C a_0^{5/2}\Biggl[\frac{1}{2.7+490(\nu-0.4654)^2}+100|\nu-0.4654|^3-
\nu a_0^{1/2}
\Bigl(\frac{1}{1.1058+6.5985|\nu-0.4736|} $$
$$+11307(\nu-0.4673)^4-258.26\nu(\nu-0.4759)^2\Bigr)-
a_0\Bigl(\frac{0.0813}{1.0
+6.9(1.0-\frac{0.4566}{\nu})+76.187\nu|\nu-0.4566|)}\Bigr)$$
\begin{eqnarray}
-0.05511\Bigl(\frac{1}{a_m^{1/2}}+\frac{0.2398}{a_m}\Bigr)\Bigl(1.9503-1.712\nu +0.583|\nu-0.452|\Bigr)\Biggr].
\label{graphiteperp3}
\end{eqnarray}
{\bf 4. For $2500\leq \lambda\leq 4000$\AA~ (UV region II)}:
$$K=C a_0^{5/2}\Biggl[\nu^{1/2}\Bigl(0.1161+0.60\nu-1.9684\nu^{1/4}(0.4-\nu)^{1/2}(\nu-0.25)\Bigr)-a_0^{1/2}\nu^2\Bigl(0.461+$$ 
$$1.7567|\nu-0.325|+180.6(\nu-0.325)^3\Bigr)-a_0\nu^3\Bigl(20664(\nu-0.3244)^4 +$$
\begin{eqnarray} 
\frac{1.3019\nu}{1.0+304.9(1.0-\frac{0.3229}{\nu})^2}\Bigr)-0.0507\Bigl(\frac{1}{a_m^{1/2}}+\frac{0.278}{a_m}\Bigr)\Biggr]
\label{graphiteperp4}
\end{eqnarray}
\subsection{\bf  Homogeneous graphite grains with refractive index $m=m_{\parallel}$ 
}
{\bf 1. For $1000\leq\lambda\leq 1120$\AA~ (FUV region I)}:
$$K=C a_{0}^{5/2}\Biggl[\nu^{3/2}\Bigl(0.3473-6.284(1-\nu)^2\Bigr)-a_0^{1/2}\nu^{3/2} 
\Bigl(
2.3047\nu-2.0047 + $$
$$ 1.38(1-\frac{0.901}{\nu})(\frac{1}{\nu}-1)^{1/2}\Bigr)-
{a_0}\Bigl(5.55(\frac{1}{\nu}-1)^{1/2}(1-\frac{0.901}{\nu})|1-\frac{0.9524}{\nu}|^{1/2}+$$
\begin{eqnarray}
\frac{0.2979}{\nu^{1/2}}-\frac{11.65}{\nu^{1/2}}(1-\nu)^2\Bigr)-\Bigl(\frac{0.046}{a_m^{1/2}}+\frac{0.011}{a_m}\Bigr)\Biggr].
\label{graphiteparl1}
\end{eqnarray}
{\bf 2. For $1120\leq\lambda\leq 1500$\AA~ (FUV region II)}:
$$K=C a_0^{5/2}\Biggl[0.06685 -0.06056\nu +0.277\nu^2-a_0^{1/2}\Bigl(0.15482\nu-0.07954\Bigr) - $$
\begin{eqnarray}
a_0\Bigl(0.75484\nu -
0.495\Bigr)-
0.018\Bigl(\frac{2}{a_m^{1/2}}+\frac{1}{a_m}\Bigr)\Biggr]  
\label{graphiteparl2}
\end{eqnarray}
{\bf 3. For $1500\leq\lambda\leq 2400$\AA~ (FUV region III)}:
$$K=Ca_0^{5/2}\Biggl[ \frac{1}{\nu^{1/2}}\Bigl(0.0864+\frac{0.79}{\nu}(\nu-0.4675)^2-7.16(\nu-0.49)^4\Bigr) -\frac{a_0^{1/2}}{\nu^{5/2}}\Bigl(0.0065+$$
$$\frac{0.06575}{\nu}(\nu-0.4922)^2 - 
1.76(\nu-0.5112)^4\Bigr)
-a_0\Bigl(0.0054 + 
\frac{0.24|\nu-0.5263|}{1.0+118(nu-0.5405)^2}\Bigr)$$
\begin{eqnarray}
-\frac{0.03913+0.07925|\nu-0.5311|}{a_m^{1/2}}-\frac{0.008-0.2116(\nu-0.5486)^2}{\nu a_m}\Biggr]
\label{graphiteparl3}
\end{eqnarray}
{\bf 4. For $2400\leq\lambda\leq 4000 $\AA~ (UV)}:
$$K =Ca_0^{5/2}\Biggl[0.1495-\frac{0.0847}{\nu}|\nu-0.3675|-a_0^{1/2}\Bigl(\nu (0.1915-0.817|\nu-0.3719|)-$$
$$33.0|\nu-0.3675|(0.4167-\nu)(\nu-0.25)^2\Bigr)-a_0\frac{0.00356}{(1+15.017|\nu-0.4031|)^2}-$$
\begin{eqnarray}
\frac{0.26}{a_m^{1/2}}\nu(1.0-1.36\nu)-\frac{0.00436+ 0.0122\nu|\nu-0.3542|}{a_m\nu}\Biggr]
\label{graphiteparl4}
\end{eqnarray}
\subsection {\bf Homogeneous silicate grains}
{\bf 1. For $1000\leq\lambda\leq 1460$\AA~ (FUV region I)}:
$$K=Ca_0^{5/2}\Biggl[0.11668-0.02745\nu + 0.17\nu^2 -a_0^{1/2}\Bigl(0.3542\nu-0.2110\Bigr) -$$
\begin{eqnarray}
-a_0\Bigl(0.6677\nu-0.3733\Bigr)+\Bigl(2a_m(1-\nu)(\nu-0.6856)-\frac{0.05}{a_m^{1/2}}-\frac{0.0062}{a_m}\Bigr)\Biggr]
\label{silicate1}
\end{eqnarray}
{\bf 2. For $1460\leq\lambda\leq 2000$\AA~ (FUV region II)}:
$$K=Ca_0^{5/2}\Biggl[0.29168\nu-0.02268-a_0^{1/2}\Bigl(0.0634\nu -0.01623\Bigr)-$$
\begin{eqnarray}
a_0\Bigl(0.64634\nu-0.3478\Bigr)-\Bigl(\frac{0.05}{a_m^{1/2}}+\frac{0.0062}{a_m}\Bigr)\Biggr]
\label{silicate2}
\end{eqnarray}
{\bf 3. For $2000\leq\lambda\leq 4000$\AA~ (UV)}:
$$K=Ca_0^{5/2}\Biggl[-0.02833+0.197\nu+\frac{0.00669}{\nu}+0.006567a_m\nu -\frac{0.0151}{a_m^2\nu}-$$
\begin{eqnarray}
-\frac{0.016417}{a_m^3}\Bigl(\frac{1-4|1-\frac{0.435}{\nu}|}{(1.0+10|1-\frac{0.435}{\nu}|)^2}\Bigr)\Biggr]
\label{silicate3}
\end{eqnarray}
In all the above formulae (\ref{graphiteperp1})-(\ref{silicate3}), $C=\pi N/10^8;$
  $a_0$, $a_m$  are in  units 
of $10^{-5} cm $; $\nu$ in units of   
   $10^{5} cm^{-1}$.
\section{Numerical comparisons}
Figure 1 shows a comparison of exact extinction curves with predictions of 
 formula for graphite grains. The $"\frac{1}{3}-\frac{2}{3}"$ approximation 
has been used in these comparisons. The three graphs presented in this figure 
are for $a_0= 0.002\mu m, 0.0035\mu m$ and $0.005\mu m$ with fixed value of     
$a_m=0.25\mu m$. It can be seen  
 that the agreement between exact results and predictions of formulas is 
excellent. Although not shown here, it has been noted that while the variation
 in $a_0$ results in bigger changes in extinction, the variation in extinction 
due to variation in $a_m$ is comparatively much weaker. \\\\
For silicate grains, the comparison of predictions from (\ref{silicate1}), 
(\ref{silicate2}) and (\ref{silicate3}) have been compared with exact results 
of computations from (\ref{extinction})  in figure 2. 
 In this figure, the three admitted values of $a_0$ are   
 $0.004\mu m~,0.005\mu m$ and $0.006\mu m$ with fixed 
$a_m=0.30\mu m$. Predictions of formula can be seen to be extremely good. Here
 also the variation in extinction with $a_m$ is much weaker in comparison to 
that with $a_0$.\\\\
To give an idea of  errors in predictions of the formulas obtained, 
we have displayed percent error details in figures 3, 4 and 5 respectively for graphite 
perpendicular, graphite parallel and silicate grain formulae. The percent error has 
been defined as:
$$Percent~ error = \frac{\bigl[K_{ext}(exact)-K_{ext}(formula)\bigr]\times 100}{K_{ext}(exact)}$$
In general, the error is within $1\%$ but for a few regions it can go upto 
about $3\%$.  The results are representative results for particular 
values of $a_0$ and $a_m$. However, It has been varified that this result is 
valid for entire range of 
$a_0$ and $a_m$ values considered in this work.
 
\section{Conclusions and discussions}
In this work, we have presented the extinction spectrum analysis for 
astronomical silicate and graphite grains in the wavelength range $\sim$ 
1000 - 4000 \AA~ which covers the FUV and UV regions. The grain size 
distribution range covered by us is in keeping with the acceptable 
compositional aspects of interstellar dust models corresponding to average 
interstellar spectra data. From this point of view, we feel that the results 
and the analytic formulae presented in this paper have application-worthiness 
in the sense that accurate estimates of extinction contributions for these 
material components in the FUV and UV regions can directly be made. Needless 
to say that a fuller and more complete account of the extinction contributions 
made by each and every material components in respect of the various dust 
models are needed for ascertaining their effectiveness in reproducing the 
average interstellar dust extinction spectrum. This necessitates further 
large -scale investigative work-outs and mathematical analyzes of similar 
nature as provided by us in this work.\\\\
However, as is mentioned in section 1, our prime motivation behind this work 
has been to analyze the extinction spectrum for a size distribution of 
spherical particles having variable refractive index ($m=m(\lambda)$). Earlier, 
the same problem with constant $m$ (at all wavelengths) was analyzed to have 
simple extinction-frequency relationships in different frequency ranges 
involving the first few lower order moments as well as some coefficients which 
seemed to be arbitrarily dependent on the refractive index. In case of 
power-law distribution $f(a)=a^{-3.5}$ all the moments ($\bar{ a},~\bar{a^2},~
\bar{a^3},~\bar{a^4}$) become simple functions of the end points $a_0,~a_m$. 
Consequently, we have looked for functional forms of extinction in terms of 
variables $a_0,~a_m$ and $\nu(1/\lambda)$ in various sub-regions as has been 
given in section 3. This analysis can also be extended to more complex 
distributions that have been used by some authors [5,6]. In the present work, 
we have restricted ourselves to UV and 
FUV regions and power-law distribution for substantive demonstration of 
workability of the formulae developed along the line of our earlier work. It 
is our wish to extend this further to cover the visible and infrared regions 
of extinction spectrum for both these grain components as well in near future.  
\clearpage
\noindent
{\Large\bf References}\\\\
$[1]$ Roy AK and Sharma SK, A simple analysis of extinction spectrum of a size 
distribution of Mie particles. J Opt A: Pure and Appl Opt 2005; 7: 675-684.\\ 
$[2]$ Mathis JS, Rumpl W and Nordsieck KH, The size distribution of 
interstellar grains. ApJ 1977; 217: 425-433.\\ 
$[3]$ Draine BT and Malhotra S, On graphite and the 2175 \AA~ extinction 
profile. ApJ 1993; 414: 632-645.\\
$[4]$ Draine BT at http://www.astro.princeton.edu/draine.\\
$[5]$ Mathis JS, Dust models with tight abundance constraints. ApJ 1996; 472: 643-655.\\
$[6]$ Weingartner JC and Draine BT. Dust grain-size distributions and extinction in the milky way, large magellanic cloud and small magellanic cloud. ApJ 2001;
 548: 296-309. 

\clearpage
\newpage
\noindent
{\Large\bf Figure captions}\\\\
Figure 1. Comparison of predictions of equations (\ref{graphiteperp1})- 
(\ref{graphiteperp4}) and (\ref{graphiteparl1})-(\ref{graphiteparl4}) in 
"$\frac{2}{3}- \frac{1}{3}$" approximation with exact 
computations from (\ref{extinction}). Solid lines are predictions and points 
are exact computations. In this figure, $a_m=0.25\mu m$ and $a_0=0.005\mu m$ 
(solid line), $a_0=0.0035 \mu m$ (large dashed line) and $a_0=0.002 \mu m$ (small 
dashed line).\\\\ 
Figure 2. Comparison of predictions of equations (\ref{silicate1}), 
(\ref{silicate2}) and  (\ref{silicate3})  with exact 
computations from (\ref{extinction}). Solid lines are predictions and points 
are exact computations. In this figure, $a_m=0.30\mu m$ and $a_0=0.006\mu m$ 
(solid line), $a_0=0.005 \mu m$ (large dashed line) and $a_0=0.004 \mu m$ (small 
dashed line).\\\\ 
Figure 3. Percent errors in equations (\ref{graphiteperp1}), 
(\ref{graphiteperp2}), (\ref{graphiteperp3}) and (\ref{graphiteperp4}) with 
respect to exact  
computations from (\ref{extinction}).  
 In this figure, $a_m=0.25\mu m$ and $a_0=0.005\mu m$. \\\\ 
Figure 4. Percent errors in equations (\ref{graphiteparl1}), 
(\ref{graphiteparl2}), (\ref{graphiteparl3}) and (\ref{graphiteparl4}) with 
respect to exact  
computations from (\ref{extinction}).  
 In this figure, $a_m=0.25\mu m$ and $a_0=0.005\mu m$.  \\\\
Figure 5. Percent errors in  equations (\ref{silicate1}), 
(\ref{silicate2}) and (\ref{silicate3})  with respect to exact 
computations from (\ref{extinction}).  
 In this figure,  $a_m=0.30\mu m$ and    
 $a_0=0.005\mu m$.  
\clearpage
\newpage
\begin{figure}[!htbp]
\centerline{\includegraphics[width=5.0in, height=10.0cm]{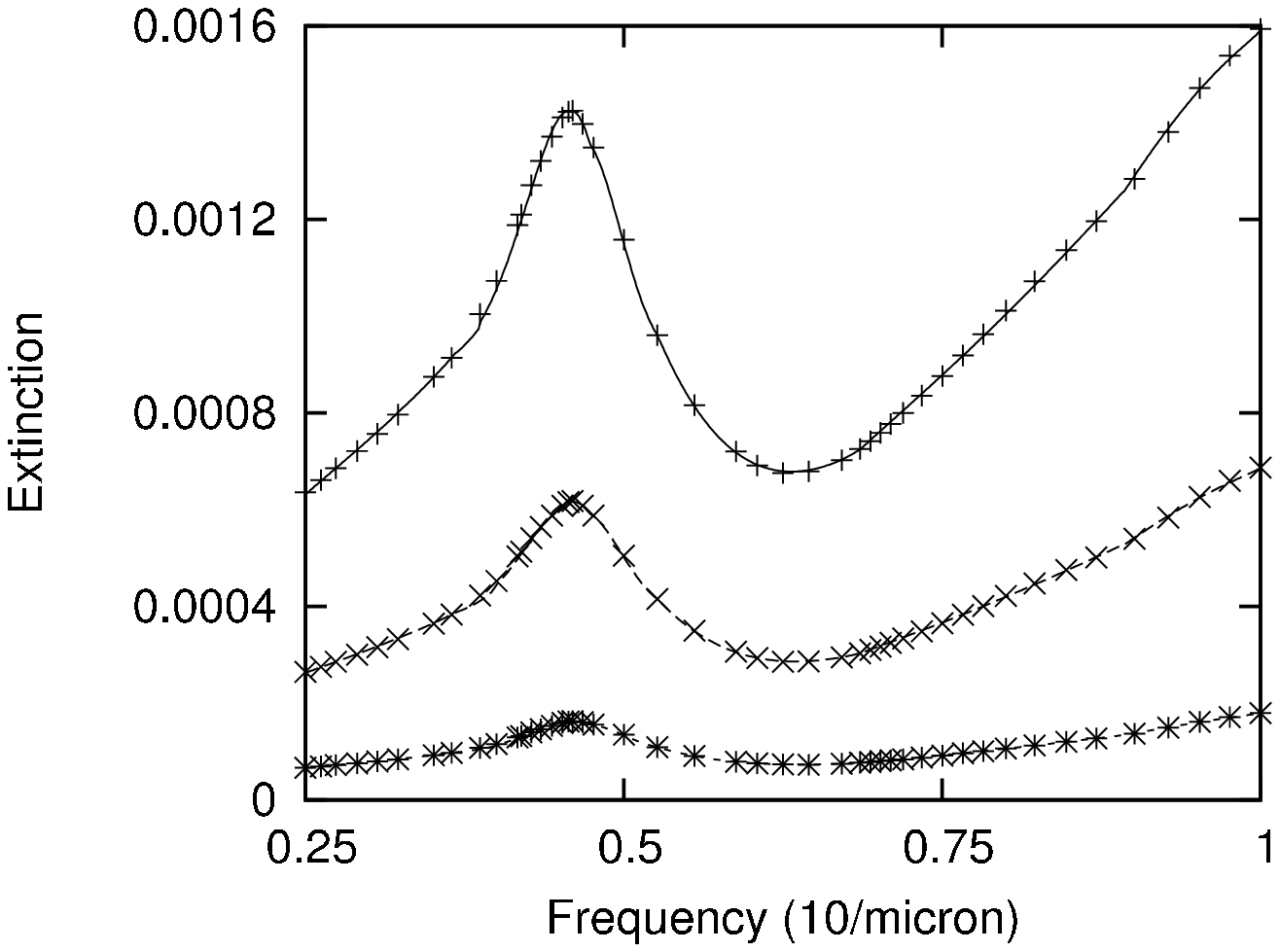}}
\vspace{0.1pc}
\caption{}
\end{figure}
\newpage
\noindent
\begin{figure}[!htbp]
\centerline{\includegraphics[width=5.0in, height=10.0cm]{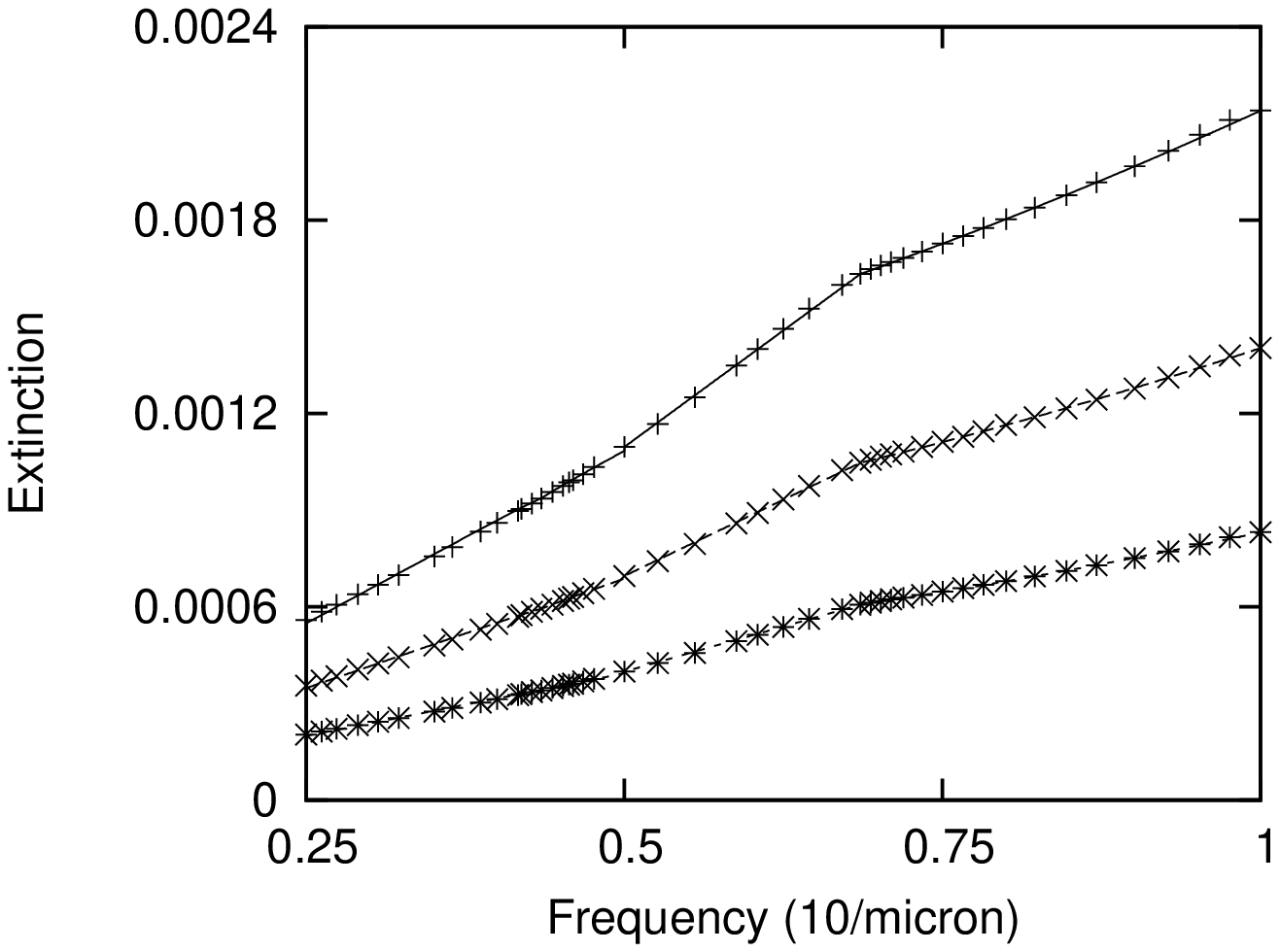}}
\vspace{0.1pc}
\caption{}
\end{figure}
\clearpage
\newpage
\begin{figure}[!htbp]
\centerline{\includegraphics[width=5.0in, height=10.0cm]{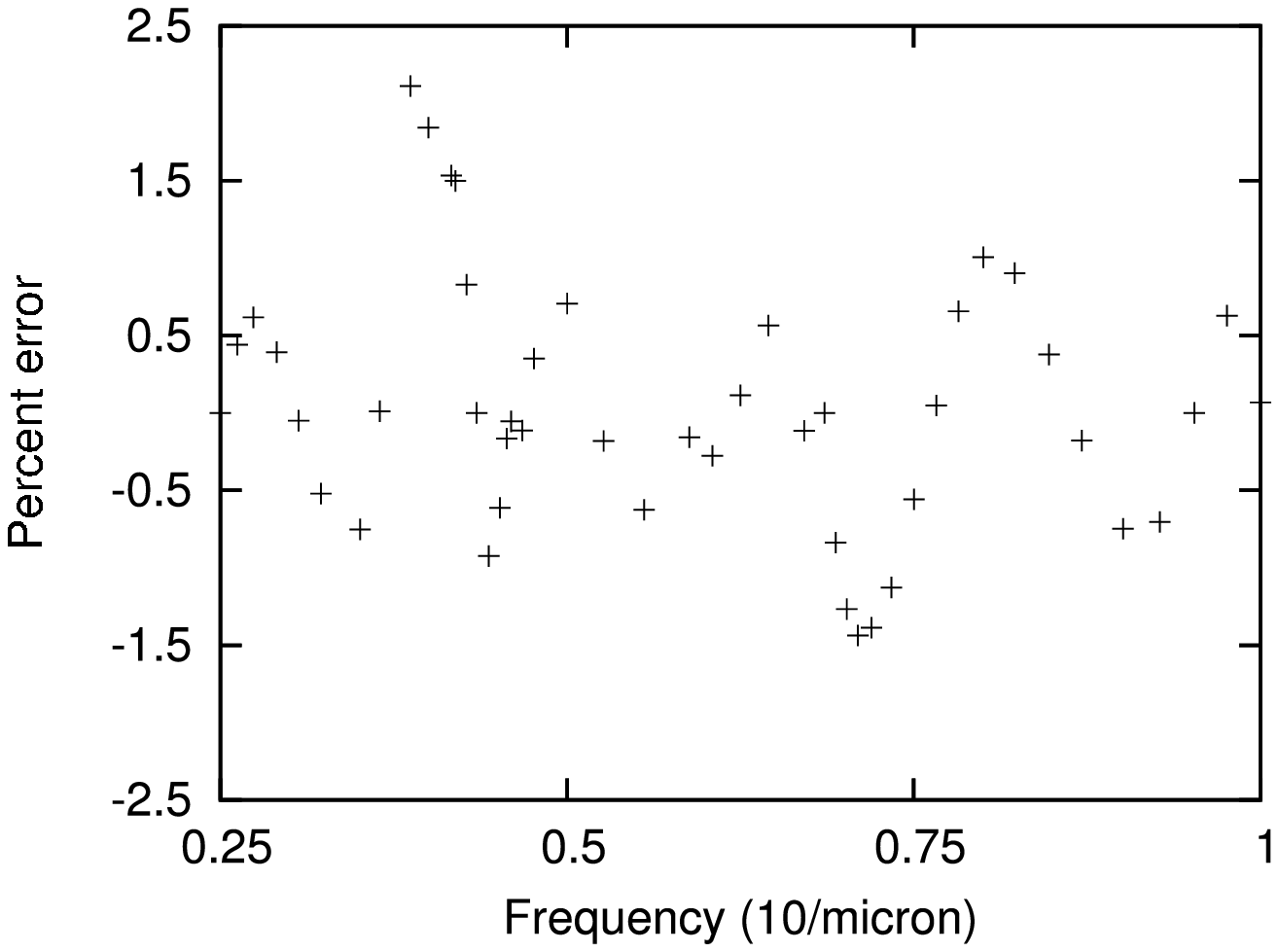}}
\vspace{0.1pc}
\caption{}
\end{figure}
\clearpage
\newpage
\noindent
\begin{figure}[!htbp]
\centerline{\includegraphics[width=5.0in, height=10.0cm]{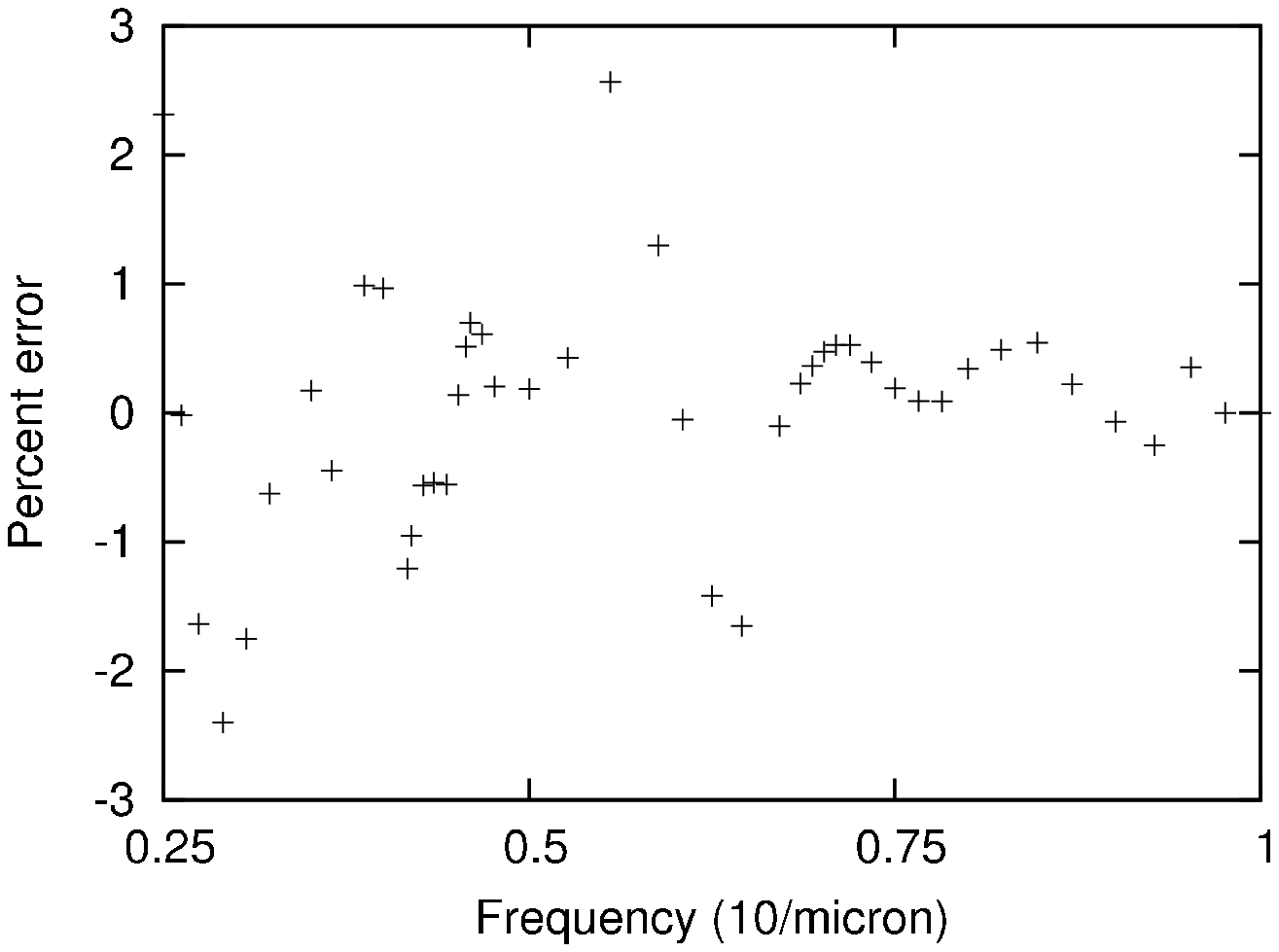}}
\vspace{0.1pc}
\caption{}
\end{figure}
\newpage
\noindent
\begin{figure}[!htbp]
\centerline{\includegraphics[width=5.0in, height=10.0cm]{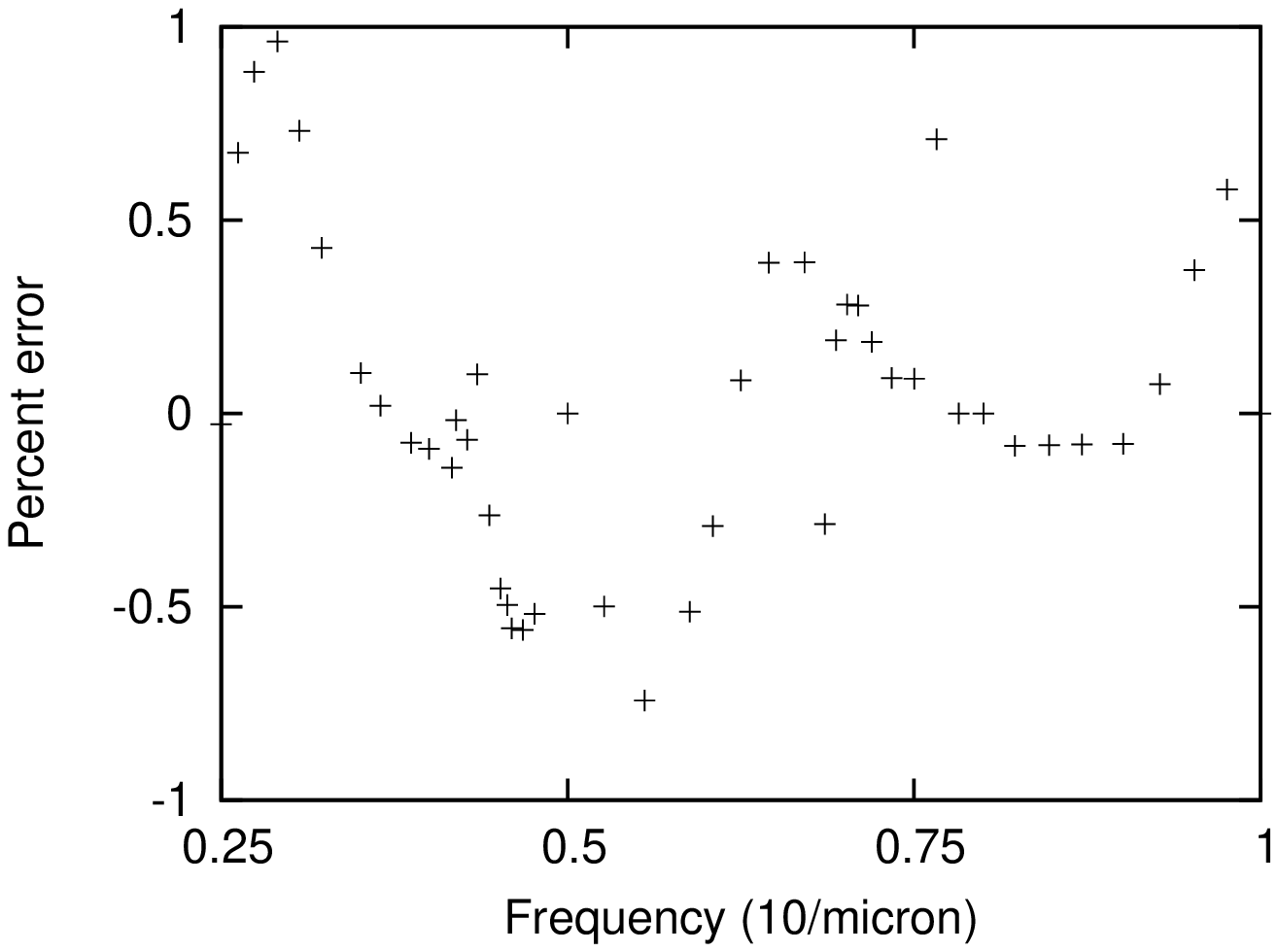}}
\vspace{0.1pc}
\caption{}
\end{figure}
\clearpage
\end{document}